\documentclass[twocolumn,prb,amsmath,amssymb]{revtex4}
\usepackage{graphicx}
\usepackage{bm}
\begin{document}

\title{Development of the tunnelling gap in disordered 2D electron system\\ with magnetic field: observation of the soft-hard gap transition}

\author{Yu.~V.~Dubrovskii$^{1}$, V.~A.~Volkov$^{2}$, L.~Eaves$^{3}$, E.~E.~Vdovin$^{1}$, 
O.~N.~Makarovskii$^{3}$,\\ J.-C.~Portal$^{4,6,7}$, M.~Henini$^{3}$ and G.~Hill$^{5}$}

 \affiliation{$^{1}$~Institute of Microelectronics Technology RAS, 142432 Chernogolovka, Russia
\\ $^{2}$~ Institute of Radioengineering and Electronics RAS, Moscow, Russia
\\ $^{3}$~ The School of Physics and Astronomy, University of Nottingham, Nottingham NG7 2RD, United Kingdom
\\ $^{4}$~Grenoble High Magnetic Field Laboratory, MPI-CNRS, BP166 38042 Grenoble Cedex 9, France
\\ $^{5}$~Department of Electrical and Electronics Engineering, University of Sheffield, Sheffield S3 3JD, United Kingdom
\\ $^{6}$~INSA, F31077 Toulouse Cedex 4, France
\\ $^{7}$~Institut Universitaire de France, 103, Boulevard Saint-Michel, 75005 Paris, France}
\begin{abstract} Magnetic field suppression of the tunneling between disordered 2D electron systems in GaAs around zero bias voltage has been studied. Magnetic field B normal to the layers induces a dip in the tunneling density of states (TDOS) centered precisely at the Fermi level, i.e. soft tunneling gap. The soft gap has a linear form with finite TDOS diminishing with B at the Fermi level. Driven by magnetic field the transition soft-hard gap has been observed, i.e. the TDOS vanishes in the finite energy window around Fermi level at $B>13$~T.
\end{abstract}

\maketitle

The problem of strongly correlated electrons is a focus of modern condensed matter physics. In part this is the problem of localized in space interacting electrons. It is well known, that as the result of electron-electron interaction in localised regime the single particle density of states (DOS) of disordered electron system tends to zero at the Fermi level \cite{1,2}. To describe that Efros and Shklovskii \cite{1} (ES) proposed in 1975 the following universal soft gaps of the DOS near the Fermi level at $T=0$ which is called Coulomb gap: parabolic gap for 3D case $G_{3D}(\varepsilon)=\alpha (\varepsilon^2 k^3)/e^6$, and linear gap $G_{2D}(\varepsilon)=\alpha' (|\varepsilon| k^2)/e^4$    
for two dimensional electron system (2DES). Here  $\varepsilon=E-\mu$, $\mu$ is a chemical potential, $k$~--~dielectric constant, $e$~--~electron charge, $\alpha$   and $\alpha'$   are numerical constants. Soft gap means that the DOS are not vanished anywhere except may be at the centre of the gap contrary to the case of a hard gap when the density of states equals to zero in the finite energy range.

Evidently, that the most direct way to observe and study the Coulomb gap is tunnelling experiment. Nevertheless, since ES prediction the main efforts were directed to study variable range hopping (VRH) conductance in disordered electron systems where in presence of Coulomb gap the Mott low, i.e. $\ln{\sigma} \propto T^{1/p}$ , where $p=4$ for 3D, and $p=3$ for 2D, is not more valid. Instead $\ln{\sigma}\propto T^{1/2}$  is expected for both dimensionality. The problem in transport measurements is difficulty to distinguish between two temperature laws, especially in 2D case. Thus most experimental studies in 2D VRH regime have reported either Mott \cite{3} or ES \cite{4} behaviour. Recently crossover of these two regimes with temperature in heterostructure based 2DES have been reported \cite{5}.

The first direct observation of the Coulomb gap by tunnelling experiment in bulk nonmetallic Si:B semiconductor was claimed only few years ago \cite{6}. Direct evidence of the expected linear form of the Coulomb gap for the 2DES  has been reported much later \cite{7} in ultrathin quench-condensed insulating Be film. These studies were carried our without magnetic field.

On the other hand, near fifteen years studies \cite{8,9,10,11,12,13,14,15} have clearly demonstrated that normal to the layers magnetic field suppresses the tunnelling of low-energy electrons between 2D electron systems or into 2D electron one from adjacent 3D metallic layer. In all cases the suppression was attributed to the formation of the gap around Fermi level in the tunnelling density of states (TDOS) due to the intra-layer Coulomb interaction. Electron-electron Coulomb interaction becomes essential in a magnetic field since in a high enough magnetic field electrons can be considered as a wave packets of the size of the magnetic length, which is less then average inter-electron distance in the layer.

Hard gaps as well as soft gaps induced by magnetic field were found in the TDOS of the 2DEGs \cite{8,9,10,11,12,13,14,15}. Only exponentially small current was detected in a normal magnetic field between 2D layers \cite{9,10} when applied voltage bias was roughly less then electrostatic energy of electron Coulomb interaction in the layers. It was associated with the wide hard gap formation in the TDOS of each 2DEG. However, studies of tunnelling between 2D and 3D electron systems \cite{12,14} revealed the soft gap developed in magnetic field in the TDOS of the 2DES with linear energy dependence. 

It should be pointed out, that the samples used in these studies were of different quality. The best ones were employed in the work \cite{9} for 2D-2D tunnelling experiments. It was bilayer 2D electron system with single layer sheet density $\approx 1.6\times 10^{11}$~cm$^{-2}$ and low temperature mobility near $3\times 10^6$~cm$^2/Vs$. In that case the authors attributed found formation of the hard gap to the strongly correlated nature of the electron systems and argued that disorder plays only subsidiary role in they experiments. In other works \cite{10,11,12,13,14,15} the samples used had about the same sheet density but measured or estimated low temperature mobility was about one order of magnitude less. The work \cite{10} in fact supported the findings presented in the paper \cite{9} on the similar samples but with lower quality. Returning to the works, where the formation of the soft gap was revealed \cite{12,14}, we note, that the slope of the gap diminished with magnetic field in both studies. Since the gap was proportional to inverse magnetic field Chain et~al. \cite{12} suggested that their data were not able be explained by a simple Coulomb gap and proposed qualitative explanation based on local Coulomb blockade effect. Contrary to them Deviatov et~al. \cite{14} considered their finding as the manifestation of the ES Coulomb gap.

This apparent inconsistence of the published results seems to be a consequence of different disorder and its influence on the TDOS in the samples of cited works. Indeed, the mobility is not a very good figure of merit for disorder especially in tunnelling experiments where the quantum lifetime rather than the transport lifetime is crucial for tunnelling characteristics \cite{16}. Of course, disorder influences both, but it could be in a different way.

In this paper we present results of our studies of tunnelling between 2DES's with much stronger disorder than it was in the samples of the preceding works \cite{9,10,11,12,13,14,15}. One of the goals is to check weather disorder soften the gap induced by magnetic field in the TDOS of the 2DES's, which looks happened since in the highest quality samples \cite{9} the formation of the hard gap was found clearly. Another goal follows from the simple picture of electron space localization in disordered 2DES by magnetic field. One can expect that in a high enough magnetic field classical ES model could be applied and linear soft gap with DOS vanishing at Fermi level should be revealed experimentally. Unfortunately, to the best of our knowledge, the problem of the TDOS in strongly disordered 2DES in a magnetic field has not been considered theoretically so far.

To form the 2DES's we used so called $\delta$-doping technique. That is the system with extremely narrow out of plane distribution of donor or acceptor impurities (ideally located in one monolayer only) in the host semiconductor. This is a full 2D analogy of the bulk doped semiconductor. The reasons for our choice are the following.
 
(1) The low temperature mobility in this kind of 2DES's with electron sheet concentration  \mbox{$(1 \div 3)\times 10^{11}$~cm$^{-2}$} is about few thousands of cm$^2/Vs$ \cite{17}. That means that in our samples the disorder is roughly two orders of magnitude more strong, than it was in the samples used in the preceding works \cite{10,11,12,13,14,15}.

(2) From equality ${2r_B/ \sqrt{N_D}}=1$ , where $N_D$ is a sheet donor concentration and $r_B$ -- Bohr radius, one can easily estimate the critical doping concentration $N_C$ corresponding approximately to an insulating regime of the two-dimensional electron gas. For the 2DES with electron concentration slightly above $N_C$ the insulating regime can be achieved varying electron concentration by means of gate electrode or by applying perpendicular to the 2DES plane magnetic field, which shrinks the electron wave functions in the simplest picture. Main studies of VRH regime \cite{4,5} revealed the ES law of conductance temperature dependence, were carried out with strongly disordered 2DES formed by $\delta$-doping.

(3) The strong disorder in this kind of 2DESs was confirmed by lateral transport studies in a magnetic field \cite{18}. Only one quantum Hall states with $\nu=2$ was revealed at low temperatures. In lower and higher magnetic fields the 2DES was in insulating state with well defined critical magnetic fields of insulator -- quantum Hall liquid -- insulator transitions.

Earlier \cite{19} we reported that normal magnetic field suppresses the tunnelling of electrons with low energy between identical 2D electron layers formed by $\delta$-doping technique. In this work we present evidence that development of the soft tunnelling gap in the TDOS of the 2DES with magnetic field was the reason of the suppression. The TDOS at the Fermi level in the centre of the soft gap is finite and decreases with magnetic field. The analysis of tunnelling spectra signals that soft gap has linear form in the vicinity of the Fermi level. Moreover we have found that the soft gap converts into a hard gap in a high enough magnetic field. The hard gap starts when the TDOS at the Fermi level is vanished. 

The MBE-grown sample was a single barrier GaAs/Al$_{0.4}$Ga$_{0.6}$As/GaAs heterostructure with a 12~nm thick barrier. The barrier was separated from the highly-doped bulk contact regions by 50~nm thick undoped GaAs spacer layers. Si~donor $\delta$-layer sheets with concentration of 
$3\times 10^{11}$ cm$^{-2}$ were located 5~nm from each side of the barrier. This is slightly above the critical concentration $N_C^*=2.4\times 10^{11}$~cm$^{-2}$ estimated from equality ${2r_B/ \sqrt{N_D}}=1$ . This estimation is consistent with earlier findings \cite{17} demonstrated that 2D electron system formed by $\delta$-doping in GaAs with electron concentration $n=2.4\times 10^{11}$~cm$^{-2}$ equal to the doping level had activated conductance behaviour.

In our experiments, electron transport along the layers does not contribute to the measured current, which flows perpendicular to the plane of the barrier. The tunnelling transparency of the main barrier is much lower than that of the spacer regions, so that most of the applied voltage is dropped across the barrier. Measurements of the Shubnikov-de-Haas (SdH) like oscillations in the tunnel current gave electron sheet concentrations approximately equal to the donor doping levels in the $\delta$-layers. Samples were fabricated into mesas of diameter 50 or 100 $\mu$m. \pagebreak The band diagram of the structure without external bias is shown in Figure~1a. Most data presented were acquired at $T=0.3$~K.

\begin{figure}[t]
\includegraphics[width=3.3 in]{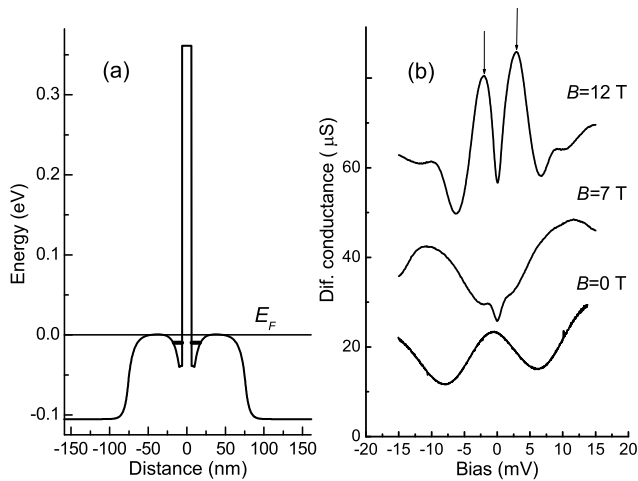}
\caption{a) Schematic band diagram of the structure without external bias. b) Differential conductance versus voltage bias in different magnetic fields. Curves (except at $B=0$~T) are shifted arbitrary in vertical direction for clarity. Arrows indicate conductance peaks used for normalization of tunnelling differential conductance (details in the text).}
\end{figure}

Figure~1b shows differential conductance in zero magnetic field, in $B=7$~T, and in $B=12$~T. A broad peak around zero bias in zero magnetic field reflects resonant tunnelling between 2D ground subbands of the 2DESs. The observation of this peak indicates also that conservation of in-plane momentum and/or small angle elastic scattering events are important for the tunnelling processes without magnetic field, despite the relatively large number of scattering centres in the 2D layers. At $B=7$~T one can see the wide minimum and narrow dip with minimum exactly at zero bias. If one supposes that in-plane momentum does not conserve in the tunnelling processes in a magnetic field, then the tunnelling differential conductance reflects convolution of the TDOS \cite{20} and appearance of the dip in the tunnelling conductance can be attributed to the formation of the gap in the TDOS by magnetic field. Since localisation is enhanced by a magnetic field, it is not surprising that the gap in the TDOS appears only at high B, when most states in the 2DES become localised. Firstly the dip in the conductance can be resolved at $B=3.5$~T. The wide minimum at $B=7$~T, when filling factor in the electron layers $\nu \approx 2$, is related with low TDOS at Fermi level located between Landau levels. With external bias increase differential conductance increases due to the transitions between Landau levels of different indexes allowed in the presence of large number of scattering centres in the layers. At $B=12$~T Fermi levels are located inside broadened ground Landau levels in both 2DESs and one sees again broad peak but with dip in the middle now, contrary to the situation in zero magnetic field.

Since tunnelling differential conductance at any constant bias oscillates, we normalized measured differential conductance $\sigma(V)$  to demonstrate how dip develops with magnetic field. The result is shown in Figure~2a. In high magnetic field we used for normalization the magnitude $\sigma_N(B)$ of differential conductance equals to the amplitude of the peaks at negative or positive bias indicated by arrow in Figure~1b. But for magnetic fields lower then $B=8$~T the peaks are disappeared. We found that peaks shifted linear with magnetic field on voltage scale in accordance with expression $V_{peak}[V]=0.215 \times B[T]-0.448$. To extend normalization procedure for tunnelling spectra acquired bellow $B=8$~T we simply used for normalization the magnitude $\sigma_N(B)$ of differential conductance at bias voltage determined by this expression. As can be seen from Figure~1b the experimental curves are slightly asymmetrical around zero voltage, therefore the results of normalization are different using amplitude of positive or negative bias peaks. The choice of negative bias sign to normalize data in Figure~2a was accidental. In what follows we use for presentation and treatment the negative bias branch of tunnelling spectra also. 

Figure~2b shows relative depth of the dip  $\sigma(0)/\sigma_N(B)$ versus magnetic field obtained by the described above procedure.

\begin{figure}
\includegraphics[width=3.4 in]{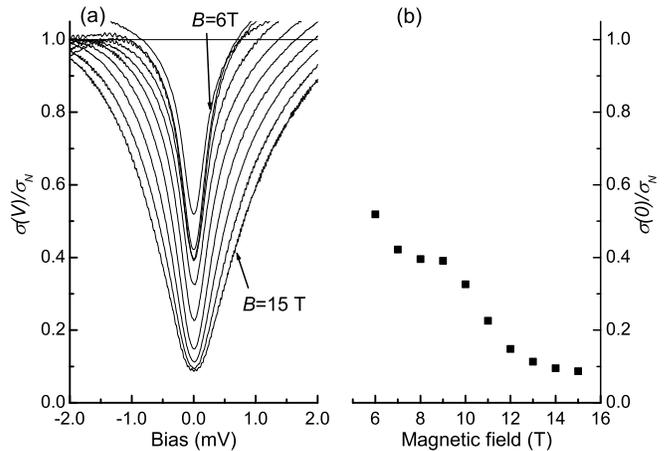}
\caption{a) Normalized tunnelling differential conductance $\sigma(V)$ in different magnetic fields. Magnetic field step is 1~T. b) Dependence of the normalized tunnelling conductance at zero bias on magnetic field.}
\end{figure}

Now we focus on the form of the dip. It was found that differential conductance in the voltage range from $V_T=3kT/e$ to $V_B$ can be well described by sublinear parabolic dependence  $\sigma=\sigma_0 +\alpha |V|- \beta V^2$, where $T=0.3$~K is the temperature of measurements, $k$ -- Boltzmann's constant. When $V_T=90~\mu V$ was conditioned by experimental set up, the voltage $V_B$ was determined from the fitting procedure so that standard deviation from the best fit curve had not exceed 0.1\%. The voltage $V_B$ increases with magnetic fields, for $B=5$~T and 15~T it is 0.4~mV and 2.1~mV accordingly. For magnetic field less then 5~T the fitting procedure was less accurate and standard deviation was few percents. Figure~3 shows the dependences of fitting parameters $\sigma_0$ and $\alpha$ on magnetic field. 

\begin{figure}
\includegraphics[width=2.9 in]{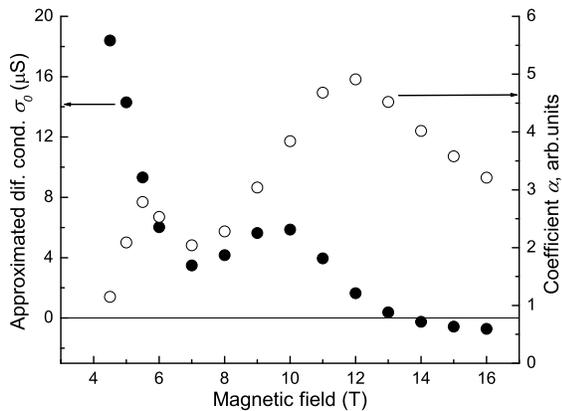}
\caption{Dependences of the fitting parameters $\sigma_0$ (filled circles) and $\alpha$ (open circles) on magnetic field, when differential conductance can be well described by \mbox{equation $\sigma = \sigma_0 + \alpha |V|-\beta V^2$ .}}
\end{figure}

The most interesting is that $\sigma_0$ becomes negative around 14~T. It would appear reasonable that negative $\sigma_0$ signals the formation of the hard gap in TDOS around Fermi level. To make the formation of the hard gap obvious, we present differential conductance around zero bias in magnetic fields 12~T, 14~T, and 16~T in the Figure~4a. The measured zero bias conductance versus magnetic field reaches minimum and tunnelling spectrum at 16~T demonstrates plato-like feature around zero bias. This transformation is more visible in the Figure~4b, which shows second derivative of the measured $I-V$ curves for the same fields. We attribute these peculiarities to the soft-hard gap transition in the TDOS of disordered 2DES's in the magnetic field about 14~T.

We can formally extract TDOS gap parameters for magnetic fields $B\leq13$ T, when $\sigma_0$ is positive. To accomplish this one should suggest that in-plane momentum does not conserve when electrons tunnel between localised states. In this condition tunnelling differential conductance described by parabolic dependence  $$\sigma=\sigma_0 +\alpha |V|- \beta V^2$$ reflects convolution of the TDOS in the 2DESs. The TDOS can be described by expression $$D(\varepsilon) \propto (\sqrt{\sigma_0}+ \frac{\alpha}{2\sqrt{\sigma_0}}|\varepsilon|+ \ldots)$$  at least in the close vicinity of the Fermi level. Figure~5 shows dependences of parameter $\sqrt{\sigma_0}$  proportional to the density of states at the Fermi level D(0), and parameter $\alpha/(2\sqrt{\sigma_0})$  proportional to the gap slope $\partial D(\varepsilon)/{\partial \varepsilon}$  on magnetic field.

\begin{figure}[t]
\includegraphics[width=3.1 in]{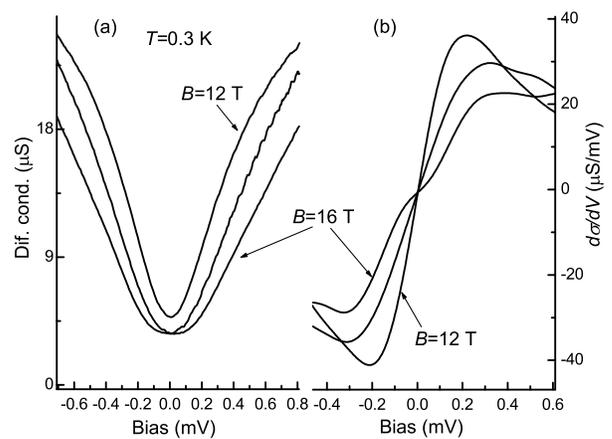}
\caption{a) Differential conductance versus voltage bias in a high magnetic field. Zero bias conductance versus magnetic field reaches minimum and tunnelling spectrum at 16~T demonstrate plato-like feature around zero bias. b) Derivative of differential conductance versus voltage bias in a high magnetic field.}
\end{figure}

If one supposes that magnetic field only decreases the localisation length of electron states, then our observation of the soft-hard gap transition contradicts to ES \cite{1,2} prediction of the existence of universal soft linear gap with vanishing TDOS at the Fermi level for strongly localized 2DES. It should be noted that magnetic field induced soft gap reported by Deviatov et.~al. \cite{14} has not vanishing density of states at the Fermi level also. On the other hand it was shown recently \cite{21} that ES~Coulomb gap in ultrathin quench-condensed insulating Be film exists only in narrow temperature range. With further temperature decrease it transforms into the hard gap. Is it related to particular properties of quench-condensed Be film, or is it the result of more sophisticated many electron phenomena, is not clear yet. While hardening of the ES Coulomb gap by electronic polarons was also considered theoretically \cite{22}. Nevertheless, we are not aware of any theoretical description in the literature of the soft gap behaviour in a magnetic field and its transformation into the hard gap.

Now we focus on the parameters of the soft gap in the TDOS of strongly disordered 2DES found here in moderate magnetic fields $B\leq 13$~T. The gap slope in the vicinity of the Fermi level (see Figure~5) is increased with magnetic field. This is opposite to what was reported earlier \cite{12, 14} for 2DES's with less disorder. Since the density of states at the Fermi level decreases with magnetic field (see Figure~5) and the gap becomes deeper, the increase of the slope with magnetic field signals that change of effective width is slow, at least much slower then in 2DES's with less disorder \cite{12, 14}. It could be reasonable because in disordered 2DES with strongly localised electrons Coulomb interaction is determined by average inter-electron spacing in a wide range of magnetic fields. The width of the gap must be of the order of Coulomb energy.

\begin{figure}
\includegraphics[width=3.4 in]{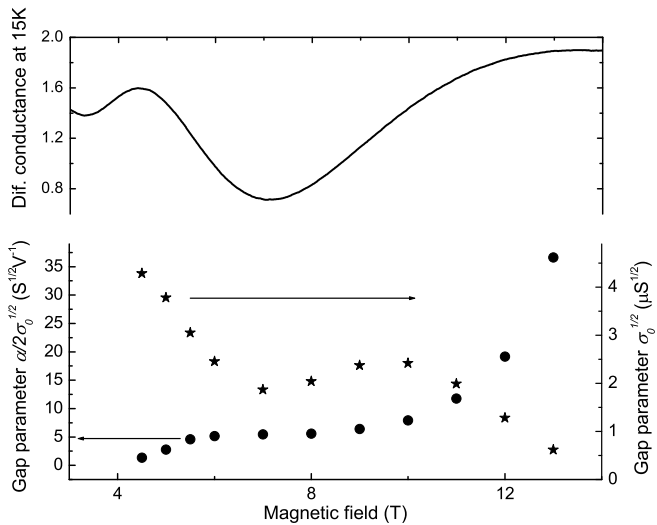}
\caption{Dependences of the TDOS parameters on magnetic field; $D(0)\propto \sqrt{\sigma_0}$ , $\partial D(\varepsilon)/\partial \varepsilon \propto \alpha/(2 \sqrt{\sigma_0})$   . Upper curve is the variation of the tunnelling differential conductance at zero bias with magnetic field measured at $T=15$~K.}
\end{figure}
 
As can be seen in the Figure~5, the TDOS in the centre of the gap has oscillating with magnetic field component. Oscillation component of $D(0)$  varies in phase with zero bias tunnelling differential conductance measured at temperature $T=15$~K (upper curve in Figure~5) when the dips in the tunnelling spectra are smeared out in all magnetic fields. It means that D(0) proportional to the unperturbative $D_0(0)$  single electron DOS in the magnetic field. Nevertheless, the relative density of state at the Fermi level $D(0)/D_0(0)$  decreases monotonically with magnetic field.

    The gap slope (Figure~5) and relative depth of the dip  $\sigma (V=0,B)/\sigma_N(B)$ (Figure~2b) have plato-like feature below $B=10$~T. Roughly above this field both parameters vary much faster with magnetic field. It looks like hardening of the gap starts there. It might be well to point out that approximately in the same field the quantum Hall liquid-insulator transition take place in the similar 2DES \cite{18}. In our structures  $\partial \sigma (V=0,B)/\partial T$ has local minima at $B=9.5$~T \cite{19} which was interpreted as the manifestation of the quantum Hall liquid-insulator transition in the tunnelling current between two near identical 2DES's. Thus we believe that our data signal about direct relation between gap evolution with magnetic field and different ground states of the 2DES's.
    
    In conclusion, we have investigated tunnelling between disordered 2DES's at temperature 0.3~K over a wide range of magnetic field applied perpendicular to the electron layers. The Coulomb interaction between electrons in the layers leads to features in the tunnelling spectrum. With increasing magnetic field, more and more electron states are localised. As the result, the narrow dip appears at zero bias, due to a soft gap in the TDOS near the Fermi level. The conversion of the soft gap into the hard gap in a high enough magnetic field was measured for the first time. 

    This work was supported by INTAS (01-2362), RFBR (04-02-16869, 02-02-22004, 02-02-17403), PICS (1577), RAS programs ``Quantum Macrophysics'', and ``Low-Dimensional Quantum Nanostructures'', FTNS program, EPSRC, and RS (UK).

\end{document}